\documentclass[reprint,amsmath,amssymb,aps]{revtex4-1}
	
\usepackage{graphicx}
\usepackage{dcolumn}		
\usepackage{bm}				
\usepackage{url}

\usepackage{xcolor}
\newcommand{\rev}[1]{\color{black}#1\color{black}}

\begin{document}

%
%
\title{Predicting rigidity and connectivity percolation in disordered particulate networks using graph neural networks}

\author{D. A. Head}
\email{d.head@leeds.ac.uk}
\affiliation{School of Computer Science, University of Leeds, Leeds LS2 9JT, United Kingdom.}.

\date{\today}

\begin{abstract}
	Graph neural networks can accurately predict the chemical properties of many molecular systems, but their suitability for large, macromolecular assemblies such as gels is unknown. Here, graph neural networks were trained and optimised for two large-scale classification problems: the rigidity of a molecular network, and the connectivity percolation status which is non-trivial to determine for systems with periodic boundaries. Models trained on lattice systems were found to achieve accuracies $>95\%$ for rigidity classification, with slightly lower scores for connectivity percolation due to the inherent class imbalance in the data. Dynamically generated off-lattice networks achieved consistently lower accuracies overall due to the correlated nature of the network geometry that was absent in the lattices. An open source tool is provided allowing usage of the highest--scoring trained models, and directions for future improved tools to surmount the challenges limiting accuracy in certain situations are discussed.
\end{abstract}

\maketitle

%
%
\section{Introduction}

The percolation transition from finite to system-spanning structures as an order parameter crosses a critical threshold has proven to be a powerful concept in statistical physics, encompassing such diverse problems as the spread of contagion and forest fires, and the reaction of polymers to form a macroscopic gel~\cite{StaufferAharony,Wierman1981,Griffiths2017,RubinsteinColby}. Closely related to this scalar transport problem is the vector propagation of forces through rigid sub-structures, leading to rigidity percolation when one such structure spans the system and bulk rigidity is realised~\cite{Maxwell1864,Calladine1978,SahimiBook,Moukarzel1995}. For models to provide practical support for such problems, they should be able to rapidly determine the connectivity and rigidity percolation status of given microscopic conformations; however, such algorithms are not always straightforward to develop, and may consume significant computational resources. While it is straightforward to identify system-spanning clusters using the classic Hoshen-Kopelman algorithm~\cite{Hoshen1976}, determining when clusters connect with themselves through periodic boundaries, and hence percolate, requires a more sophisticated, graph-based approach which is not trivial to implement or parallelize~\cite{Livraghi2021}. Rigidity percolation is even more challenging. The pebble game provides an integer-based algorithm to determine the rigidity of two-dimensional networks with independent bond constraints~\cite{Jacobs1995}, but more general problems require the calculation of system-wide response to an applied load, which can require sophisticated solvers to be developed and optimized~\cite{Head2003b,Wilhelm2003,Vermeulen2017}.

Machine learning (ML) has emerged as a powerful tool when applying classification and regression tasks to a broad range of physical problems~\cite{Carleo2019,vanMastrigt2022,Lu2024,Buehler2024,Raissi2019,Karniadakis2021,Penwarden2022,Saadat2024}. Applications to the percolation transition include unsupervised ML based on clustering algorithms employed to localise the connectivity transition~\cite{Yu2020,Zhang2022,Mimar2022}. Additionally, Bayo \emph{et al.} leveraged supervised ML in the form of convolutional neural networks, a class of model that has been substantially developed for image classification, to predict connectivity percolation in two-dimensions~\cite{Bayo2022}. However, the more challenging problem of predicting the rigidity of a system from the geometry of its constituent parts has not yet been addressed by ML. Moreover, current approaches are most immediately suited to classification problems and may not be easily extended to regression tasks. This suggests that novel approaches will need to be explored for future ML models to predict continuous properties of a material of given microstructure and composition.

Here, we develop and evaluate an ML scheme for the classification of both rigidity and connectivity percolation of a given set of particles, and the bonds that connect them. The model belongs to the class of graph neural networks (GNNs), so-called because they take sets of nodes (particles) and edges (bonds) as inputs, and can be trained to generate predictions for either single nodes in one large graph (\emph{e.g.}, a social network), or of entire graphs themselves~\cite{Reiser2022,Gurnani2023}. In the latter capacity, GNNs have been extensively applied to predict various properties of molecules~\cite{Park2022} and aid the discovery of novel therapeutic drugs~\cite{Wang2023}. This suggests GNN-based ML models applied to molecular gels have the potential for regression analysis of \emph{e.g.} bulk mechanical properties. However, it is first sensible to evaluate their efficacy on simpler classification tasks. Here we consider two related problems: lattices of springs (bonds) with a fraction of springs randomly removed; and off-lattice particles that dynamically cross\-link into gels. All systems are two-dimensional with periodic boundaries
	\rev{unless otherwise stated. We find that both systems can be reliably trained for the classification of both connectivity and rigidity percolation, even when applied to system sizes an order of magnitude different to that on which the model was trained, albeit with reduced accuracy. However, }
	the accuracy of predictions for the off-lattice data are significantly lower than for the lattice data. We argue this is due to the spatial correlations inherent in the dynamical simulations but absent in the bond-diluted lattices, and suggest means to improve accuracy without requiring the need to generate unrealistically large data sets for training.

%
%
\section{Methodology}

\subsection{Data Generation}
\label{s:dataGeneration}

Lattice data was generated using code previously developed to determine the viscoelastic properties of two-dimensional immersed spring networks~\cite{Head2019,Head2022}. In brief, regular triangular lattices of $L\times L$ nodes connected by Hookean springs were generated in which a fraction $1-p$ of springs were randomly removed. Percolation of clusters of nodes connected by springs was determined using the algorithm of Livraghi \emph{et al.} that incorporates periodic boundaries~\cite{Livraghi2021}, returning a percolation dimension of 0, 1 or 2. Rigidity was determined by calculating the viscoelastic response under simple shear at an arbitrarily low frequency, with purely local frictional drag at lattice nodes, and rigidity identified with a non-zero storage modulus. Prior to rigidity determination, node positions were randomly perturbed to avoid possible artefacts induced by colinear springs~\cite{Calladine1978,Vermeulen2017}, with natural spring lengths set to inter-node distances after this perturbation to eliminate internal stresses. For details of network generation and storage modulus calculation see~\cite{Head2019,Head2022}

Off-lattice data was generated by Brownian dynamics code devised for simulating collagen gelation~\cite{Head2016}. In contrast to the lattice model, thermal fluctuations are now included, \emph{i.e.}, the microscopic thermal energy scale $k_{\rm B}T>0$, so connectivity and rigidity percolation are expected to occur at similar densities. $N$ particles of diameter $d$ were placed in an $L\times L$ periodic box, and diffused with a coefficient $D=k_{\rm B}T/\gamma$ in terms of the drag coefficient~$\gamma$. Particles repelled \emph{via} a soft repulsion that allows small overlaps; particles that overlap become connected by a permanent cross\-link, \emph{i.e.}, a spring of stiffness $k$ and natural length $d$. After a predetermined time~$t_{\rm max}$, the connectivity percolation dimension is determined as above, then further cross\-linking is suppressed and the system is sheared through Lees-Edwards boundary conditions~\cite{AllenTildesley} at a frequency~$\omega$ for 5 cycles. The shear modulus is extracted from the final cycle, and the network is identified as rigid when this average exceeds a pre\-determined threshold. Suitable values for $t_{\rm max}$, $\omega$, and the threshold, were determined from preliminary runs. These are summarised in Table~\ref{t:beadSpringParams}.

\begin{table}
	\begin{center}
		\renewcommand{\arraystretch}{1.25}
		\begin{tabular}{c@{\:\:}|@{\:\:}ccc}
			$N$ & $t_{\rm max}/t_{\rm diff}$ & $\omega t^{\rm d}_{\rm diff}$ & $G_{\rm thresh}^{\prime}(\omega)d^{2}/k_{\rm B}T$
			\\
			\hline
			$10^{2}$ & 1000 & $10^{-2}$ & 0.15
			\\
			$10^{3}$ & 1000 & $10^{-2}$ & 0.1
			\\
			$10^{4}$ & 1500 & $3\times10^{-3}$ & 0.05
		\end{tabular}
	\end{center}
	\caption{The total time $t_{\rm g}$, shear frequency $\omega$, and shear modulus threshold $G^{\prime}_{\rm thresh}$, for $N$ particles in the off-lattice model. $t_{\rm max}$ and $\omega$ are both normalised by the time $t^{\rm d}_{\rm diff}$ for a particle to diffuse its own diameter, and the (two-dimensional) shear modulus threshold is normalised by $k_{\rm b}T/d^{2}$.
	}
	\label{t:beadSpringParams}
\end{table}

All networks were stored in the portable \texttt{.json} file format containing fields for the node indices (unique integers), edges (pairs of node indices), and the classification outputs of the connectivity percolation dimension (integer 0, 1 or 2) and the rigidity (integer 0 or 1 for non-rigid or rigid, respectively). The node positions and edge lengths were also stored to test for their potential impact on training and accuracy. These files represent a general format that could be generated from the outputs of any general particle-based simulation package. These generic data were then converted to the PyTorch Geometric class \texttt{InMemoryDataset}~\cite{PyG,PyTorch}.

\subsection{Machine Learning Model}
\label{s:model}

The networks generated in Sec.~\ref{s:dataGeneration} were represented as undirected graphs $\mathcal{G}=(\mathcal{V},\mathcal{E})$ consisting of nodes (particles) $i\in \mathcal{V}$, and edges (bonds) between nodes $(i,j)\in\mathcal{E}$, with $(i,j)$ and unordered pair of node indices. Each node was initially assigned a feature vector that could in principle encode molecular properties~\cite{Reiser2022}, but for the percolation problems considered here, which are expected to depend primarily on the connectivity encoded in $\mathcal{E}$, nodes were initially assigned the uniform scalar feature $h_{i}=1$. Graph neural networks for graph classification update node features based on a message-passing protocol that proceeds in two stages~\cite{Wang2023}. First, messages $m_{i}^{(k+1)}$ at level $k+1$ are constructed from the features at the previous level $k$, of the node $i$ itself, and of all nodes $j$ in its neighbourhood $N(i)=\{j\in\mathcal{V}\,|\,(i,j)\in\mathcal{E}\}$,
\[
	m_{i}^{(k+1)} = \sum_{j\in N(i)} \phi_{\rm mess}^{(k)}
	\left(
		h_{i}^{(k)},h_{j}^{(k)},e_{ij}
	\right),
\]
	the message functions $\phi_{\rm mess}^{(k)}$ to be determined. The edge feature $e_{ij}$ could in principle encode bond features such as stiffness or length. Node features are then updated as
\[
	h_{i}^{(k+1)} = \phi_{\rm upd}^{(k)}
	\left(
		h_{i}^{k},m_{i}^{k}
	\right),
\]
	with the update functions $\phi_{\rm upd}^{(k)}$ again to be determined. After $n_{\rm H}$ such message passing layers, a linear readout layer maps all node features to the required outputs.

	It remains how to determine $\phi_{\rm mess}^{(k)}$, $\phi_{\rm upd}^{(k)}$, and the weight matrix in the readout layer. To respect translational invariance, the spectral convolution graph neural network of Defferrard \emph{et al.} was employed that maps between layers using the graph Laplacian approximated by a polynomial filter of order $K-1$~\cite{Defferrard2016}. The optimised version of this algorithm is implemented in PyG as \texttt{ChebConv} layers, parameterised by $K$ and the number of output channels per layer $n_{\rm C}$~\cite{PyG}. The parameters in each of the $n_{\rm H}$ \texttt{ChebConv} layers and the readout matrix are trained to minimise the loss function, here chosen to be the cross-entropy loss commonly used for classification problems~\cite{AggarwalBook}, including a weighting proportional to the size of each class~\cite{PyTorch}.
	
\subsection{Training Protocol}
\label{s:training}

Unless otherwise stated, each data set was separated into training, validation, and testing data sets in the ratio 4:1:1 respectively. The training data was used to update model parameters so as to reduce the error (loss) using Adam optimization with a constant learning rate $\ell_{r}$ and batch size $n_{\rm batch}$~\cite{HaykinBook}. The data was randomly shuffled before each epoch (\emph{i.e.}, each sweep through the training set). After each epoch, the validation data was used to evaluate the loss independently of the training data. Training terminated if (i)~a pre-specified maximum number of epochs was reached; (ii)~the loss evaluated on the validation data averaged over 10 epochs increased by over 10\% of its lowest value (suggesting over-fitting), or (iii)~the evaluation loss did not improve for 50 epochs of training.

Cross-validation was performed by splitting the training and evaluation data into $k$ groups, cyclically selecting $k-1$ groups for training and the remaining 1 for evaluation. The overall accuracy was averaged over all $k$ `folds' on the unseen test data, with uncertainty estimated by the standard error of the mean. Preliminary tests for each data set with the smallest number of samples (largest system size) was used to estimate a suitable number of folds and maximum number of epochs as shown in Fig.~S1 of the supplementary material~\cite{SI}. Each of the 5 remaining meta\-parameters were then varied to determine suitable ranges, as shown in Fig.~S2. Suitable $\ell_{r}$ and $n_{\rm batch}$ were fixed based on these results, and the remaining meta\-parameters $n_{\rm H}$, $n_{\rm C}$ and $K$ were systematically varied over the ranges given as shown in Tables S1 to S4, to determine the best-performing model for data set and system size, separately for rigidity and connectivity classification problems.

\subsection{Evaluation Metrics}
\label{s:metrics}

Trained models are always evaluated on the unseen testing data set, which has the same distribution amongst classes as the training and testing sets. The accuracy is simply defined as the fraction of predictions made by the model that matched the labelled data. The confusion matrix $C_{ij}$ is defined as the number of samples predicted by the model to be in class~$i$ that actually belonged to class~$j$. Two common class-specific metrics are then the precision, defined as the fraction of predictions for a given class that were correct, and the recall, defined as the fraction of all actual data points belonging to that class for which the model prediction was correct. It will sometimes be convenient to combine these into the single $f1$ score, being the harmonic mean of both; thus, the $f1$ score for class $i$ is (no summation over repeated $i$ indices)
\[
	\left(f1_{i}\right)^{-1}
	=
	\left(\frac{C_{ii}}{\sum_{k=1}^{n_{\rm class}}C_{ik}}\right)^{-1}
	+
	\left(\frac{C_{ii}}{\sum_{k=1}^{n_{\rm class}}C_{ki}}\right)^{-1}
	\quad.
\]
Note that the accuracy can be defined in terms of $C_{ij}$ as the trace divided by the sum of all matrix elements.

%
%
%

%
%
\section{Results}
\label{s:results}

Rigidity percolation is a binary classification problem in which a system can be either rigid or non--rigid, whereas the connectivity indentification algorithm classifies networks as percolating in 0, 1 or 2 dimensions for two-dimensional systems, \emph{i.e.}, ternary classification. The binary problem is considered first.

\subsection{Rigidity Percolation}

\label{s:rigidity}

For spring lattices, values of the dilution parameter $p$ were chosen in the range 0.60 to 0.70 inclusive with an increment of 0.01, with the number of networks per $p$ dependent on the lattice size $L$. The resulting data sets were well balanced, with approximately 53\% non-rigid and 47\% rigid systems, consistent with the known percolation threshold for infinite systems of $p_{\rm c}\approx0.66$~\cite{SahimiBook}. After preliminary trials varying the meta-parameters as described in Sec.~\ref{s:training}, the number of hidden layers $n_{\rm H}$, the number of output channels per layer $n_{\rm C}$, and the Chebyshev filter $K$, were systematically varied for each system size $L$. The final accuracies are given in Table~S1 of the supplementary materials~\cite{SI}. The higher Chebyshev filter $K=4$ consistently provided the highest accuracies, with no clear preference for $n_{\rm H}$ and $n_{\rm C}$. Accuracies were high, increasing with system size from around 93\% for $L=20$ to 97\% for $L=100$, despite the reduction in the number of training samples with increasing~$L$.

\begin{figure}[htbp]
	\begin{center}
		\includegraphics[width=8cm]{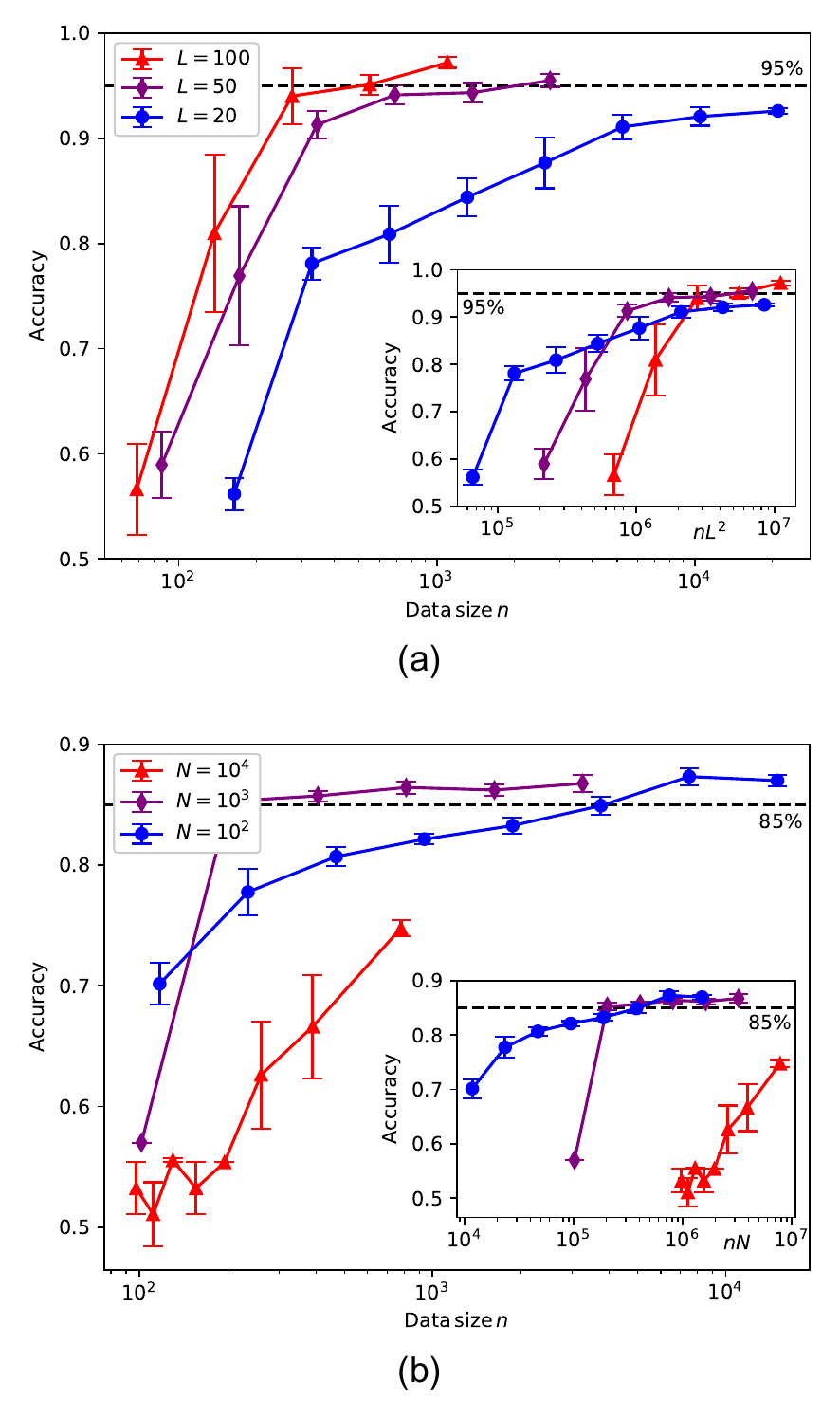}
		\caption{The accuracy of rigidity percolation classification on (a) $L\times L$ lattices and (b) $N$-particle off-lattice simulations, for varying data sample sizes $n$. Error bars are the standard error over $k=5$ independent folds of the train and test sets. \emph{(Insets)} Same data scaled to the system volume, $nL^{2}$ or $nN$.}
		\label{f:varDataSize}
	\end{center}
\end{figure}

To evaluate the dependency on the size of the training data sets, training and evaluation was repeated for restricted data sets in which a fraction of training points were removed while maintaining class balance. The accuracies are plotted in Fig.~\ref{f:varDataSize}(a) and demonstrate the expected trend of increasing accuracy for larger data sets, with the indication that greater accuracies could be attained for the larger systems $L\geq50$ if more samples were available. Given that samples for smaller systems can be generated more rapidly than large systems, it is more meaningful to plot against the total system volume rather than the number of samples. To this end, the inset to the figure shows the accuracies plotted against $nL^{2}$, and demonstrates a partial collapse of the $L=50$ and $L=100$ data at the highest $nL^{2}$, with the $L=50$ reaching the collapsed region before $L=100$.

To generate the off-lattice data for simulation sizes of $N=10^{2}$, $10^{3}$ and $10^{4}$ particles, the box dimensions were varied to generate data spanning from low density systems that never formed rigid networks, to high density systems that always formed a rigid gel, with as many points in between as resources allowed. Class balance was approximately realised, with roughly 50\% of data points being classified as rigid as per the criterion of Sec.~\ref{s:dataGeneration}. As with the lattice data, model hyper\-parameters were systematically varied to identify an optimal combination, as summarised in Table~S3 of the supplementary materials~\cite{SI}, and demonstrates that $K=n_{\rm H}=4$ and $n_{\rm C}=20$ was the preferred choice for all $N$. However, the achieved accuracies were lower than the lattice data, being always under 90\%. Varying the training sample size exhibited the same trends as for the lattice data as shown in Fig.~\ref{f:varDataSize}(b), but noticeably the small system sizes $N=10^{2}$ and $N=10^{3}$ approach a plateau accuracy while the largest system $N=10^{4}$ was still increasing in accuracy with the largest sample size realised. Plotting the same data scaled by the volume (which is $\propto N$) confirms that $N=10^{4}$ is an outlier, with only the two smaller systems demonstrating any collapse. Possible reasons for this are discussed in Sec.~\ref{s:discussion}.

\begin{figure}[htbp]
	\begin{center}
		\includegraphics[width=8cm]{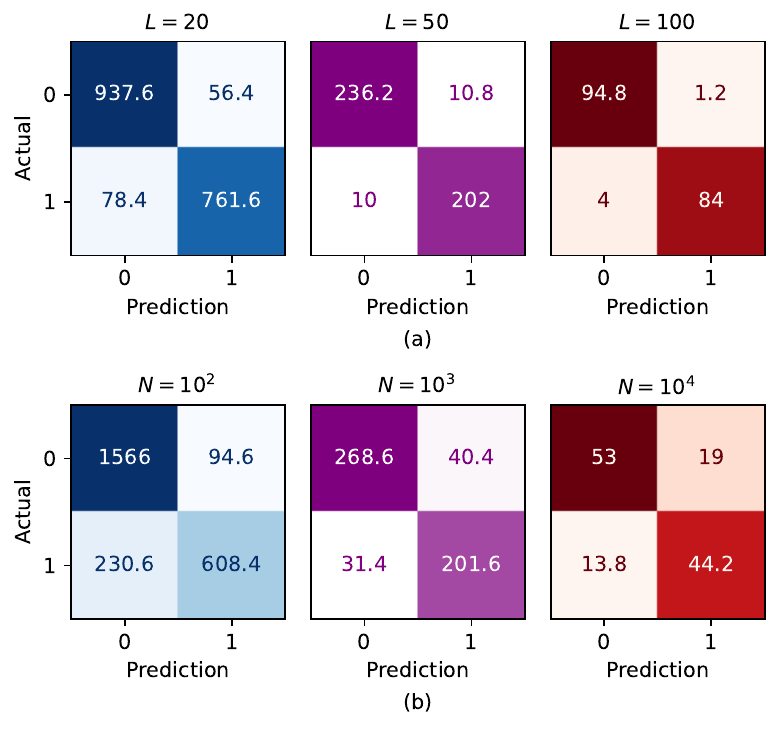}
	\end{center}
	\caption{
		Rigidity percolation classification confusion matrices for (a) lattice and (b) off-lattice models, showing numbers of testing data samples with prediction (columns) against actual (rows) classification, averaged over $k=5$ folds of training. The system size increases from left to right as annotated. Class labels 0 and 1 denote non-rigid and rigid respectively.
	}
	\label{f:confMat_rig}
\end{figure}

Insight into model bias can be improved by inspection of the confusion matrix defined in Sec.~\ref{s:metrics}. The confusion matrices for rigidity classification for parameters that attained the highest accuracy score are presented in Fig.~\ref{f:confMat_rig} for both lattice and off-lattice models. The matrices are approximately symmetrical, suggesting no strong bias among incorrect predictions. A possible exception is the smallest off-lattice system $N=10^{2}$, for which the model appears to have a greater tendency to classify rigid systems as non-rigid than \emph{vice versa}, but there is no clear trend across all models.

\subsection{Connectivity Percolation}

\label{s:connectivity}

Connectivity percolation requires non-trivial algorithms for systems with periodic boundaries. The graph-based algorithm of Livraghi \emph{et al.}~\cite{Livraghi2021} determines the connectivity percolation dimension to be an integer $d_{\rm conn}$ in the range $[0,d]$ with $d$ the system dimension, which in 2D means there are 3 possible classes. However, and crucially different to rigidity percolation, the middle class $d_{\rm conn}=1$ is a minority class with typically far fewer data points that $d_{\rm conn}=0$ (not percolated) or 2 (fully percolated). Moreover, since the percolation transition occurs over an increasingly narrow range as the system size increases~\cite{StaufferAharony,SahimiBook}, $d_{\rm conn}=1$ becomes an ever smaller class. Class imbalance must therefore be addressed.

The simplest strategy to restore class balance is to oversample; that is, to repeat each instance of a data point in the minority class $o_{1}$ times in the training set. To first determine suitable meta\-parameters for the lattice data, the integer $o_{1}$ for which the number of points in class $d_{\rm conn}=1$ with oversampling, $o_{1}n_{1}$, most closely matched the mean of the majority classes, $\frac{1}{2}\left(n_{0}+n_{2}\right)$, was determined. Accuracies varying meta\-parameters with this overweighting is given in Table~S2 of the supplementary material~\cite{SI}, and although the overall accuracies are lower than for the rigidity classification problem, the trends are similar. Note that, as the connectivity transition is distinct from rigidity for lattice models, a separate data set comprising dilution parameters $p$ in the range 0.2 to 0.4 inclusive, in steps of 0.01 as before, was generated and used for training and evaluation.

\begin{figure}[htbp]
	\begin{center}
		\includegraphics[width=9cm]{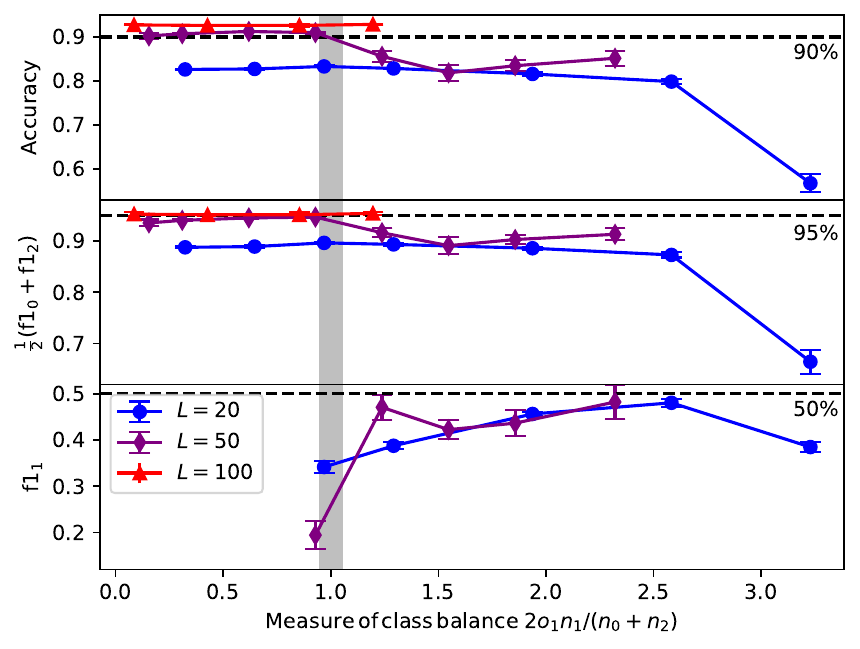}
		\caption{Accuracy (top panel), the mean of the $f1$ scores for the two majority classes (middle panel), and the $f1$ score for the minority class (bottom panel), for connectivity percolation classification of lattice data with the minority class oversampled by a factor $o_{1}$. The shared horizontal axis is an approximate measure of class balance, given $n_{i}$ data points in class $i$. The vertical shaded line denotes when class balance with oversampling is approximately achieved.
		}
		\label{f:varOverSamp}
	\end{center}
\end{figure}

To determine the effectiveness of overweighting, the accuracy, the mean $f1$ score for the majority classes (see Sec.~\ref{s:metrics}), and the $f1$ score for the minority class, are presented in Fig.~\ref{f:varOverSamp} for varying $o_{1}$. The quantity on the horizontal axis, defined in terms of the number of samples in each of the three classes, $n_{0}$, $n_{1}$ and $n_{2}$, is chosen to approach 1 when all classes are balanced during training (note $n_{0}\approx n_{2}$ for this data). It is immediately apparent that the accuracy barely exceeds 90\%, with no suggestion of improvement as the degree of oversampling increases. The $f1$ scores for the majority classes follow the same trends as the accuracy, with values just reaching 95\%. The $f1$ score for the minority class, by contrast, never significantly exceeds 50\%, and moreover cannot even be defined for low values of $o_{1}$, when the model never predicts class 1 outputs for at least 1 training fold. The conclusion is that oversampling is not an effective means to address class imbalance for this class of problem.

For the off-lattice data generated by simulations that included thermal motion of the particles, the connectivity and rigidity transitions approximately coincide, depending on the thresholds used to identify each transition. Therefore the same data sets were used for both problems. Class imbalance for connectivity was more pronounced than for the lattice data, and in addition, $n_{0}$ and $n_{2}$ were also quite different from each other. It was therefore necessary to introduce two overweighting factors, $o_{1}$ and $o_{2}$, to approach class balance. As shown in Table~S4 of the supplementary material~\cite{SI}, similar accuracies were attained as for the rigidity data, although with greater noise making it difficult to discern clear trends. As no significant changes were observed, $o_{1}$ and $o_{2}$ were not systematically varied.

\begin{figure}[htbp]
	\begin{center}
		\includegraphics[width=9cm]{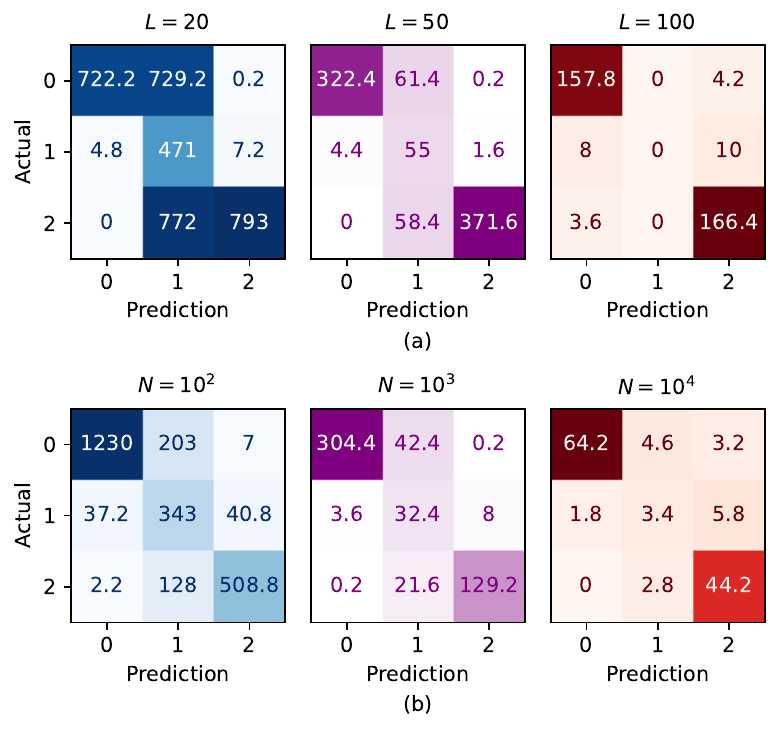}
	\end{center}
	\caption{
		Confusion matrices for connectivity percolation classification, for (a) lattice and (b) off-lattice models, showing numbers of evaluation data samples with prediction (columns) against actual (rows) classification, averaged over 5 folds. The classes 0, 1 and 2 refer to the connectivity dimension $d_{\rm conn}$.
	}
	\label{f:confMat_conn}
\end{figure}

The confusion matrices for connectivity percolation classification are given in Fig.~\ref{f:confMat_conn} for the models that achieved the highest accuracies. For the small and middle system sizes for lattice and off-lattice models, there is a clear tendency for the models to incorrectly predict connectivity in one dimension only, $d_{\rm conn}=1$, for systems that actually had $d_{\rm conn}=0$, 1 or 2. This bias appears to be diminished for the largest systems, but the reduced numbers of samples with $d_{\rm conn}=1$ makes it difficult to discern actual trends. We conclude that the low accuracy scores for connectivity percolation is primarily due the models trained with over-sampling, over-predicting the minority class $d_{\rm conn}=1$.

\subsection{Model and Data Variations}
\label{s:variations}

The following variations of the GNN and the data sets were also considered: (i)~using non-spectral (\emph{i.e.}, non-convolutional) graph neural networks; (ii)~including scalar edge weights in the data sets, equal to the distance between the connected nodes; (iii)~including the positions of the nodes; and (iv)~combining both (ii) and (iii). For the off-lattice data, an additional variation was also considered: (v)~only including the largest connected cluster in the data, which can be rapidly determined numerically, and could in future applications be added to both the data generation and prediction workflows. The accuracies for these model variations for lattice and off-lattice data are presented in Tables S5 and S6 respectively of the supplementary information~\cite{SI}. 

It is clear from the data that the non-convolution model performs significantly worse than the convolutional layers used previously, with no clear evidence that any meaningful training was achieved. This may be because convolutional neural networks are constructed to impose the translational invariance that percolation must obey, but non-convolutional models would need to be trained to respect such invariance, requiring far more training data. It may also be that, as percolation is fundamentally a long-range phenomenon, GNNs could be prone to the known problem of `over-smoothing' during many passes of message passing~\cite{Reiser2022}, which the spectral nature of convolutional neural networks may mitigate.

Augmenting the data to include lengths of edges between nodes, node positions, or both, did not alter the degree of training achieved to any measurable degree. Since this extra information expands storage requirements and resources required to train, such information should not be included for classification problems, although we cannot rule out any benefit for regression. Using only the maximum cluster also showed no significant changes in accuracy, but in this case there was a reduction in storage requirements by $\sim 50\%$ near the percolation point, and a similar reduction in training times. Given the ease with which this feature can be implemented, future applications should incorporate this filter.

\begin{figure}[htbp]
	\begin{center}
		\includegraphics[width=8cm]{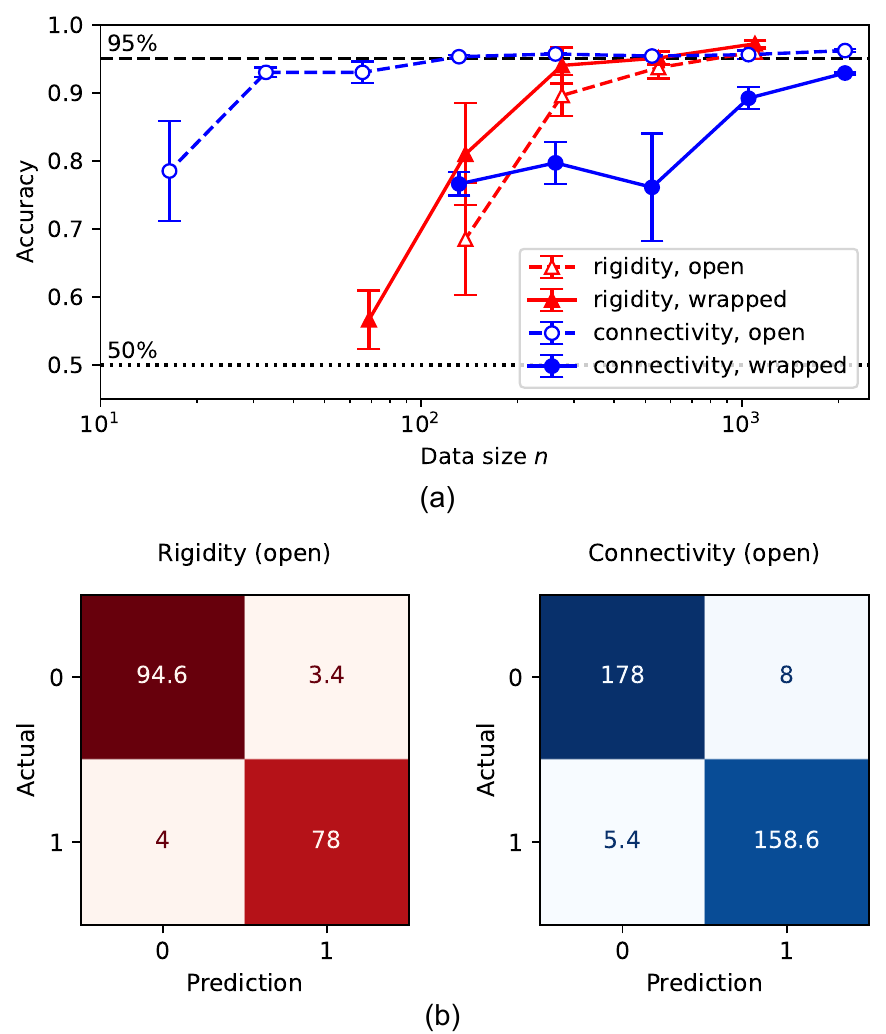}
	\end{center}
	\caption{
		\rev{Lattice models ($L=100$) with open boundaries in one dimension. (a) Variation of accuracy with data size for connectivity and rigidity percolation classification. Fully wrapped boundaries as considered elsewhere are shown for comparison. (b) Confusion matrices for rigidity and classification matrices for open boundaries, for which connectivity becomes a binary classification problem.}
	}
	\label{f:openBdy_final}
\end{figure}

%
%
\rev{Finally, to probe the role of boundary conditions, the lattice model was modified so that the network was connected through one periodic boundary only; the transverse direction had open boundaries. This means connectivity percolation can only happen in 0 or 1 dimensions and becomes a binary classification problem. This change also necessitated the use of extensional (rather than simple) shear to test for rigidity, but the generation of the raw data was otherwise as for the fully wrapped boundaries considered previously. Fig.~\ref{f:openBdy_final} shows the variation of accuracy with sample size and the confusion matrices, and shows that the accuracy for rigidity percolation follows a similar trend regardless of boundary, with values within errors. Connectivity percolation maintains higher accuracy for open boundaries than wrapped, unsurprising since the class imbalance issue has been removed. We conclude that the internal representations of GNNs are sufficiently general to be able to accurately predict percolation classification for boundary conditions typically included in simulations for bulk material properties.}

\subsection{Sensitivity to System Size}

Although message passing GNNs can be applied to graphs of any size, it does not follow that models trained on one system size will work equally well on smaller or larger systems. To test the sensitivity of models to the system size they were trained on, the accuracy for all combinations of model and data system sizes were calculated, for both rigidity and connectivity problems, and for both the lattice and off-lattice data. The results are presented in Fig.~\ref{f:sensitivity}. For off-lattice data, there is a clear decrease in accuracy when applying a model to systems that are much larger, or much smaller, than the data they were trained on. As evident from the tables, accuracies decrease by approximately 10-20\% for an order of magnitude difference in system size $N$. Thus, to achieve reasonable accuracies for off-lattice systems, it will be necessary to train models for a range of system sizes and to select the closest to the test system.

The lattice data exhibits an unusual trend in which accuracies for system sizes larger than that for which the model were trained tend to increase. As evident in the figure, the lowest accuracies for both rigidity classification arise for models trained on $L=100$ data applied to $L=20$ systems, but the highest accuracies are found for the $L=100$ data sets, with no significant variations between the models. The reason for this is not clear, but may be related to the preparation of the lattice systems, in which bonds were randomly diluted without generated spatial correlations in the system structure. As discussed later in Sec.~\ref{s:discussion}, this could enhance the effects of self-averaging, in which the model is effectively being applied to multiple small systems within one large system.

\begin{figure}
		\renewcommand{\arraystretch}{1.15}	
		\begin{tabular}{c@{\:\:}|@{\:\:}ccc}
			\multicolumn{4}{l}{Rigidity, lattice}\\
			 & $L=20$ & $L=50$ & $L=100$ 
			\\
			\hline
			$L=20$ & 0.926(3) & 0.964(1) & 0.976(4)
			\\
			$L=50$ & 0.89(1)  & 0.955(4) & 0.979(4)
			\\
			$L=100$ & 0.86(2) & 0.940(7) & 0.972(5)
		\end{tabular}
		\vspace{0.3cm}
		\\
		\begin{tabular}{r@{\:\:}|@{\:\:}ccc}
			\multicolumn{4}{l}{Connectivity, lattice}\\
			 & $L=20$ & $L=50$ & $L=100$ 
			\\
			\hline
			$L=20$ & 0.833(2) & 0.895(4) & 0.930(4)
			\\
			$L=50$ & 0.817(1) & 0.910(2) & 0.933(3)
			\\
			$L=100$ & 0.798(4) & 0.902(1) & 0.929(2)
		\end{tabular}
		\vspace{0.3cm}
		\\
		\begin{tabular}{r@{\:\:}|@{\:\:}ccc}
			\multicolumn{4}{l}{Rigidity, off-lattice}\\
			& $N=10^{2}$ & $N=10^{3}$ & $N=10^{4}$ 
			\\
			\hline
			$N=10^{2}$ & 0.870(4) & 0.76(2)  & 0.732(7)
			\\
			$N=10^{3}$ & 0.72(2)  & 0.868(7) & 0.58(3)
			\\
			$N=10^{4}$ & 0.754(8) & 0.79(1)  & 0.748(7)
		\end{tabular}
		\vspace{0.3cm}
		\\
		\begin{tabular}{r@{\:\:}|@{\:\:}ccc}
			\multicolumn{4}{l}{Connectivity, off-lattice}\\
			& $N=10^{2}$ & $N=10^{3}$ & $N=10^{4}$ 
			\\
			\hline
			$N=10^{2}$ & 0.83(1)  & 0.75(1)  & 0.47(2) 
			\\
			$N=10^{3}$ & 0.727(2) & 0.860(7) & 0.61(1)
			\\
			$N=10^{4}$ & 0.661(3) & 0.80(2)  & 0.860(7)
		\end{tabular}
	\caption{
		Series of tables showing the accuracy of models trained on data of one size (rows), applied to data sets of all sizes (columns), for rigidity and connectivity percolation on lattice and off-lattice systems as indicated. Model hyper\-parameters were those that gave the highest accuracy on the same data set as training (diagonal entries). The values in brackets are the standard errors in the last digit. 
	}
	\label{f:sensitivity}
\end{figure}

%
%
\section{Discussion}
\label{s:discussion}

The results presented here demonstrate that graph neural networks can be used to predict the properties of macromolecular gels, as demonstrated for the rigidity and connectivity percolation status of particulate systems in periodic geometries. The trained models that achieved the highest accuracies for all of the systems considered here are available from~\cite{repo}, including scripts demonstrating how they can be applied to networks generated by any procedure or simulation package, allowing these to be freely used for applications. However, it is also clear from Sec.~\ref{s:results} that further refinement will be required to train models achieving accuracies $>90\%$ for off-lattice models, and also connectivity percolation for all classes of system. The primary issues encountered and possible means to address them are discussed below.

For connectivity percolation, the key problem encountered was the class imbalance when trying to classify percolation in all dimensions up to the spatial dimension~$d$, as intermediate connectivity dimensions $1\leq d_{\rm conn}\leq d-1$ were significantly under-sampled when increasing system density in regular increments. Oversampling the minority classes was shown to not be an affective means to address this issue, as covered in Sec.~\ref{s:connectivity}. As there is no obvious way to generate new data points by interpolating between existing ones for this problem, other strategies should be explored. For instance, enhanced sampling near to the percolation transition to increase the sample size for intermediate points, or to omit such intermediate cases from consideration in applications for which they are not of interest.

The accuracy of predicting rigidity percolation for lattice systems was found to be high, with over 95\% achieved for the larger systems. By contrast, the off-lattice problem was unable to reach even 90\%, although systems of $N=10^{4}$ particles exhibited accuracies that were still rapidly increasing for the largest data set considered. However, the data set for $N=10^{4}$ was already large, with $n=780$ points, so simply increasing the training data would require a substantial commitment of resources and may not be practical. This data scarcity can be addressed by physics-informed approaches that augment training data with constraints obeying the known governing equations~\cite{Karniadakis2021}, but current approaches are not immediately transferable to molecular systems. Multi-fidelity neural networks that combine small numbers of high-fidelity data with large numbers of low-fidelity data have proven to be successful in various fields~\cite{Penwarden2022,Saadat2024}, but cannot be immediately applied to GNNs as current network layers do not admit inputs that are a mixture of graphs and another model's outputs, which is required by this approach. Given that regression analysis is likely to require even more data than classification, this issue should be the focus of future work.

\begin{figure}[htbp]
	\centerline{
		\includegraphics[width=8cm]{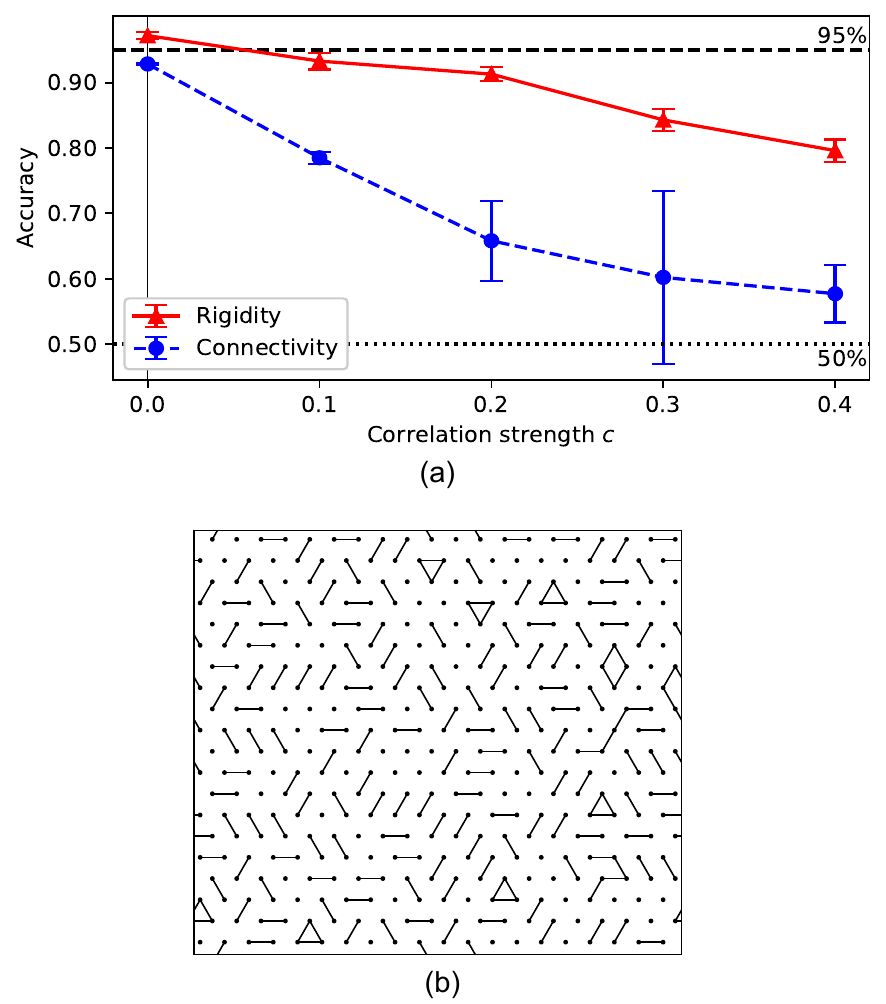}
	}
	\caption{
		\rev{Effect of correlations in lattice data. (a) Accuracy of training for $L=100$ lattices generated using the first model of Zhang \emph{et al.}, in which $c$ denotes the strength of correlations~\cite{Zhang2019}. (b) }
		Example of applying the trained model for $L=20$ rigidity percolation classification to identify the backbone, \emph{i.e.}, the minimum set of bonds required for the system to remain rigid. An $L=20$ lattice with all bonds present ($p=1$) was generated and bonds removed at random when such removal was predicted by the trained model to leave the system rigid. This clearly fails to generate the backbone, demonstrating that the model trained on uncorrelated lattices cannot be applied to correlated systems.
	}
	\label{f:corrnLattices}
\end{figure}

%
%
\rev{As for why the off-lattice data achieved lower training accuracies, it is hypothesised this is due to the inherent correlations induced by their dynamic generation, as opposed to the lattice models which were generated by bond dilution and hence  lack spatial correlations. Uncorrelated systems may act like multiple smaller, independent systems during training, increasing the effective sample size and resulting in the higher accuracies observed. To test this hypothesis, correlated lattices were generated using the first model of Zhang \emph{et al.} in~\cite{Zhang2019}, in which the strength and range of correlations were controlled by the parameter~$c$, with $c=0$ corresponding to uncorrelated lattices. As shown in Fig.~\ref{f:corrnLattices}(a), the accuracy for rigidity percolation prediction decreases from around 95\% to 80\% as $c$ increases, consistent with the reduction in accuracy between lattice and off-lattice models presented earlier. To test this hypothesis further, a scenario was devised in which a model trained on lattices without correlations was applied to a problem that requires correlations. Specifically, we attempted to identify the load-bearing backbone~\cite{Moukarzel1995} by starting from a fully occupied ($p=1$) $L=20$ lattice, and removing randomly-selected bonds if the trained model classified the system as still being rigid after the bond was removed, until no such bonds could be removed. This should leave a single rigid cluster spanning the system, but as shown in Fig.~\ref{f:corrnLattices}(b), the trained model instead produces what is clearly a non-rigid system, consisting of small, disconnected clusters. We conclude that spatial correlations are a key feature that need to be included in training data for the models to be accurately applied to correlated systems.}

\rev{In terms of performance, timing runs comparing the speed of  rigidity classification by the trained ML model compared to the pebble game algorithm~\cite{Jacobs1995} show a clear advantage to the former, taking on average 2.5(1) ms rather than 0.81(1) secs for $L=100$ lattices. Connectivity was much closer, with the algorithm of Livraghi \emph{et al.}~\cite{Livraghi2021} actually running slightly faster at 1.31(1) ms compare to 1.87(1) ms, again for comparable systems. Therefore practical benefits of this approach are currently limited to rigidity classification. However, it is expected that future work training an ML model to predict the time-dependent viscoelastic response, which currently requires extensive simulations, should drastically widen the performance gap between ML predictions and procedural calculations.}

Extending this approach to continuously--varying quantities -- \emph{i.e.}, regression rather than classification -- would expand the use of such networks beyond the small molecules to which they are currently applied~\cite{Park2022,Wang2023}. Such a rapid prediction tool could be used as part of an inverse problem workflow to determine the microscopic configuration that produces the desired measurable properties, as has already been developed for the CREASE tool that estimates the composition of nanoparticles generating a given scattering profile~\cite{Heil2022}.
	\rev{CREASE is only able to achieve this by using a pre-trained machine learning model, as generating the many data points mapping microscopic configurations to experimental measurements required during optimization would be prohibitively slow using traditional, brute-force methods. It is expected that the development of a comparable tool capable of reliably predicting the viscoelastic response of model gel systems would be of benefit to the modelling community, removing the need to implement suitable boundary conditions and run systematic frequency sweeps in time-consuming production runs, accelerating modelling support to gel-based systems. Moreover, this development would also open up the possibility of a CREASE-like model to estimate the microscopic properties of a gel given experimental viscoelastic data, lowering the need to perform scattering experiments and the lengthy, costly process of applying for access to high-end scattering facilities as is currently typical for many classes of gel.}


%
%

%
%
%
%

%
%
\nocite{*}		

\bibliography{gelPercGNN,gelPercGNN_URLs.bib}

\begin{thebibliography}{43}
\bibitem [{SI()}]{SI}%
  \BibitemOpen
  \href@noop {} {}\bibinfo {howpublished} {URL for SI will be
  added}\BibitemShut {NoStop}%
\end{thebibliography}

\end{document}